\documentclass[11pt]{article}
\usepackage{multicol}
\usepackage{moriond}
\hypersetup{colorlinks=false}

\def\Journal#1#2#3#4{{#1} {\bf #2}, #3 (#4)}

\def\be{\begin{equation}}
\def\ee{\end{equation}}
\def\bea{\begin{eqnarray}}
\def\eea{\end{eqnarray}}

\def\degr{$^{\circ}$}

\newcommand\g{\ensuremath{\gamma}}%

\newcommand{\Photo}{\includegraphics[height=0mm]{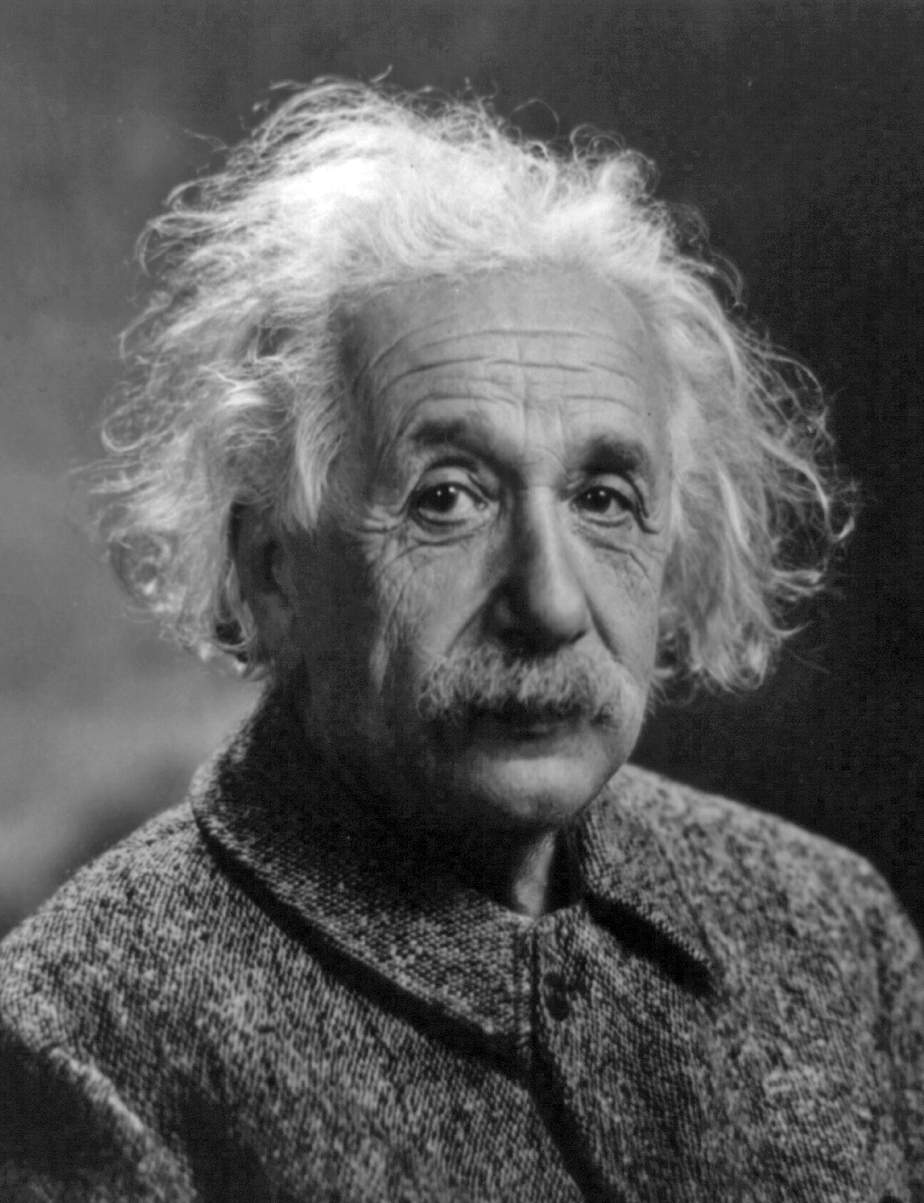}}

\begin{document}
\vspace*{3.5cm}
\title{SUPERNOVA REMNANTS INTERACTING WITH MOLECULAR CLOUDS AS SEEN WITH H.E.S.S.}

\author{ D. FERNANDEZ $^{1}$, M. DALTON$^{2}$, P.EGER$^{3}$, H. LAFFON$^{2}$, J.MEHAULT$^{2}$, S. OHM$^{4,5}$, I. OYA$^{6}$, M. RENAUD$^{1}$ FOR THE H.E.S.S. COLLABORATION }

\address{ $^{1}$Laboratoire Univers et Particules de Montpellier, Montpellier, France \\ $^{2}$Centre d'\'Etudes Nucl\'eaires de Bordeaux Gradignan, Gradignan, France  \\ $^{3}$ Physikalisches Institut, Erlangen, Germany \\ $^{4}$Department of Physics and Astronomy, The University of Leicester, Leicester, UK \\$^{5}$ School of Physics \& Astronomy, Leeds, UK \\ $^{6}$Institut f\"ur Physik, Berlin, Germany}

\maketitle\abstracts{
About 30 Galactic supernova remnants (SNRs) are thought to be physically associated with molecular clouds (MCs). These systems are prime \g-ray source candidates as the accelerated particles from shock fronts collide with the surrounding high-density medium thus emitting gamma-rays through hadronic interactions. However only a handful of such interacting SNRs are detected at TeV energies. We report the current status of the High Energy Stereoscopic System (H.E.S.S.) observations towards these SNR-MC systems, with a particular emphasis on the latest results.}

\section{Introduction}

Since the 1930s' \cite{bz}, supernova remnants (SNRs) remain the most probable sources of Galactic cosmic rays (CRs). Through the diffusive shock acceleration and subsequent magnetic field amplification mechanisms, particles can be accelerated up to the knee at the SNR shock surface. However despite several decades of multi-wavelength observations no compelling {\it direct} evidence in favor of efficient CR acceleration in SNRs has been observed yet.
The interaction of protons and nuclei accelerated at the SNR shock with the interstellar/circumstellar medium (ISM/CSM) is accompanied by the emission of high/very-high energy (HE/VHE) \g-rays resulting of the decay of neutral pions. Such hadronic interactions are tracers of the amount of protons accelerated at the SNR shock. Hence SNRs interacting with dense interstellar material such as molecular clouds (MCs) are prime targets to test the SNR paradigm. Inverse Compton scattering of accelerated leptons off the different seed photon fields or non-thermal bremsstrahlung on ambiant nuclei can also give rise to \g-ray emission. These three mechanisms lead to different photon spectral shape in the GeV-TeV energy range. One can identify the \g-ray production process by using joint observations from imaging atmospheric Cherenkov telescopes (IACTs) such as the H.E.S.S., MAGIC, VERITAS experiments as well as the Fermi-LAT and AGILE telescopes. 

The H.E.S.S. array detects \g-rays above an energy threshold of $\sim$\,100\,GeV. The primary particle direction and energy are reconstructed with an energy resolution of $\sim$\,15\% and an angular resolution of $\sim$\,0.1\degr. 
About 50 H.E.S.S. sources are coincident with a Galactic SNR (from the Green's catalogue), among which 3 are firmly associated with a SNR-MC system (W51C \cite{w51H,w51M}, W49B \cite{w49H}, W28 \cite{w28H}) and 5 with the SNR shell emission itself (Vela Jr \cite{velaH}, HESS J1731--347 \cite{j1731H}, RCW 86 \cite{rcwH}, RX J1713.7--3946 \cite{j1713H}, SN 1006 \cite{sn1006H}). The other sources are either associated with a pulsar wind nebula (PWN) emission or without any clear counterpart. 1720\,MHz OH masers are reliable indicators of a physical interaction between a SNR and a MC but other tracers like molecular (CO, CS...) line broadening or specific infrared line emissions can also be used to probe SNR-MC shocks (see B. Jiang's list of Galactic SNRs interacting with MCs at \url{http://astronomy.nju.edu.cn/~bjiang/SNR_MC.htm}).

In the following, the association of recently detected H.E.S.S. sources with SNR-MC interactions will be discussed in the light of their multi-wavelength counterparts and their spectral energy distribution modeling. Different SNR types detected in the \g-ray domain will be compared in order to emphasize some trends between isolated and MC-interacting SNRs.  

\section{ H.E.S.S. sources possibly associated with a SNR-MC shock}

\subsection{H.E.S.S. sources coincident with a SNR, a MC and OH masers}
HESS J1804--216 \cite{survey,Aje} is a bright \g-ray source with an extension of 22' and coincident with the massive star forming region W30. This complex contains several HII regions, the evolved SNR G8.7--0.1, the pulsar PSR J1803--2137 and the PWN G8.40+0.15. The presence of a MC and the detection of an OH (1720 MHz) maser \cite{HY} on the eastern edge of the remnant confirm the interaction of the SNR with the MC. However the TeV emission does not match with any known object:  the angular distance from the SNR G8.7--0.1 and from the PSR is of about $\sim$\,0.2-0.3\degr\ and the size of the coincident PWN\,(2') is much smaller than the TeV emission. While the required energy budgets do not discard any of the SNR-MC shock and PWN scenarios, such offsets make the association unclear.

HESS J1714--385 \cite{ctbH} is coincident with the SNR G348.5+0.1 (also called CTB 37A) which is interacting with three molecular clouds as shown by the detection of several 1720\,MHz OH masers in different locations in the SNR \cite{frail}. A slightly extended non-thermal X-ray source CXOU J171419.8--383023 which could be a PWN is detected in the north-western part of the remnant. However no associated pulsar is detected. On the one hand a PSR-PWN scenario is not ruled out by the estimated spin-down luminosity of a potential pulsar powering the nebula. On the other hand the derived CR energetics in the SNR-MC scenario are reasonable. Despite the slight extent of the TeV source, no detailed morphological study can be carried out for such a weak TeV source. Therefore, two possible scenarios remain to explain the HESS J1714--385 emission.

\subsection{H.E.S.S. sources coincident with a SNR and a MC}
The two TeV sources HESS J1834--087 and HESS J1640--465, both coincident with known SNRs and discovered in the H.E.S.S. Galactic Plane survey \cite{survey} have been followed up until 2012. The analysis of a larger amount of data led to more detailed morphological and spectral studies and resulted in the discovery of two new H.E.S.S. sources also coincident with SNRs: HESS J1832--093 nearby HESS J1834--087 and HESS J1641--463 close to HESS J1640--465.

HESS J1834--087 \cite{w41H} is a bright and extended VHE emission coincident with the SNR W41. The morphological analysis of the source results in the identification of two components: a TeV point-like component coincident with the central compact object (CCO) CXOU J183434.9--084443 \cite{w41X} which does not show any pulsation, and a surrounding TeV extended component with an extension of $\sim$\,0.17\degr. In addition to detecting a CCO in the center of the remnant, X-ray observations revealed a PWN candidate in the form of non-thermal diffuse emission surrounding the X-ray compact source. The TeV source is marginally coincident with $^{12}$CO \cite{w41CO12} and $^{13}$CO \cite{w41CO13} line emissions. The analysis of 47 months Fermi-LAT data results in the detection of an extended GeV source coincident with HESS J1834--087, with a power law spectrum of index $\sim$\,2.15. The two TeV components are well described by power laws with softer spectral indices ($\Gamma\ge2.6$).
A PWN scenario alone can not explain the combined spectrum of the extended GeV and TeV components. A SNR-MC scenario would imply a spectral break at $\sim$\,100\,GeV to accomodate it. A break at a few GeV is observed in the spectra of several SNRs interacting with MCs (W51C \cite{w51F}, W49B \cite{w49F}, W28 \cite{w28F}, IC443 \cite{ic4V,ic4F}, W44 \cite{w44F}) and can be explained by the damping of Alfv\'en waves due to the presence of neutrals in the dense surrounding medium \cite{malk}. However it would be the first detection of a spectral break at so high energies. Assuming such a break, the combination of a PWN and SNR-MC scenarios could explain the Fermi-LAT and H.E.S.S. spectra.

HESS J1832--093 \cite{laff} is a point-like source located 0.5\degr\ from W41 and spatially coincident with the western radio rim of the SNR G22.7--0.2. Nor the age, neither the distance of this SNR are known. An X-ray point-like source is detected 1' away from the H.E.S.S. source and no GeV counterpart is observed \cite{2fgl}. From $^{13}$CO observation \cite{w41CO13} two clouds are detected towards the TeV emission and could be interacting with G22.7--0.2. The CRs energy budget resulting from such a SNR-MC scenario is reasonable however no tracer of a SNR-MC shock has been detected so far.

HESS J1640--465 is an extended \g-ray source coincident with the northern part of the G338.3--0.0 SNR shell. This distant SNR ($\sim$10\,kpc) is coincident with the southern part of a giant HII region G338.4+0.1 and is located nearby the stellar cluster Mc81. Fermi-LAT observations revealed a GeV counterpart to HESS J1640--465. A compact and extended X-ray sources~\cite{Lem,Sla} are detected close to the center of the remnant and are compatible with the VHE emission but no pulsation has been detected. The latter source was  suggested to be a potential pulsar and HESS J1640--465 was first interpreted as the associated PWN. The new H.E.S.S. results reveal that the TeV spectrum ($\Gamma$=2.15) connects smoothly with the Fermi-LAT power-law spectrum of index $\sim$\,2.3. This differs significantly from the previous HE spectral energy distribution and suggest that at least part of the TeV emission is likely of hadronic origin. The estimated product of total energy in protons and average target gas density from 1\,GeV to 20\,TeV is $W_{p} \cdot n_{H} \approx 4 \cdot10^{52} (\frac{d}{10\,\mathrm{kpc}})\, \mathrm{erg\,cm^{-3}}$.  

HESS J1641--463 \cite{oya} is coincident with the SNR G338.5+0.1 and located $\sim$\,0.25\degr\ from HESS J1640--465. This SNR is at $\sim$11\,kpc \cite{Kot} and the giant HII region G338.4+0.1 lying between the two SNRs G338.5+0.1 and G338.3--0.0 seems to connect them. No other counterpart is found to be compatible with HESS J1641--463. The spectral analysis reveals a hard power-law shape ($\Gamma \sim$2), which explains why the distinction between the two TeV sources HESS J1640--465 and HESS J1641--463 becomes more evident above a few TeVs. 

\section{Features of isolated and MC interacting SNRs}

\begin{figure}
\centering
\includegraphics[width=\columnwidth]{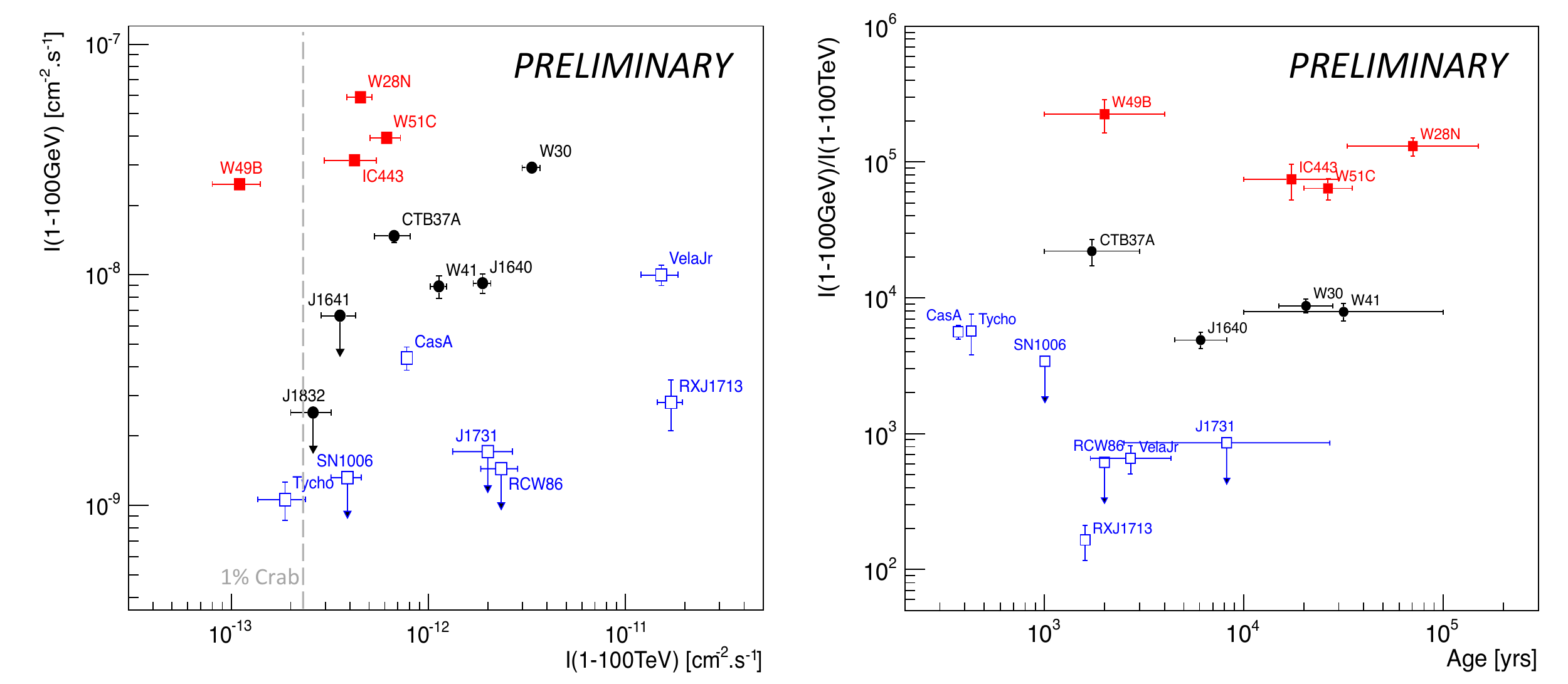}
\caption{\textit{Left:} Integrated flux on the 1-100\,GeV energy range versus integrated flux on the 1-100\,TeV energy range for SNRs interacting with MCs (full red squares), isolated SNRs (empty blue squares) and SNRs possibly interacting with MCs (black dots). The vertical dashed line represent 1\% of the Crab flux. \textit{Right:} Ratio of the  1-100\,GeV over 1-100\,TeV integrated fluxes versus the SNR age (same legend as for the left panel).} 
\label{fig:plot}
\end{figure}

Fig.\ref{fig:plot} shows the integrated 1-100\,GeV flux versus the integrated 1-100\,TeV flux (left) and the integrated flux ratio $I_{1-100\,GeV}/I_{1-100\,TeV}$ versus the SNR age (right) for different SNR types. HE-VHE sources associated with SNR-MC systems (W51C, W28N, W49B, IC443) have an integrated GeV flux much higher than the integrated TeV flux: $I_{GeV}/I_{TeV} = 5\cdot10^{4}-3\cdot10^{5}$. This confirms that SNRs interacting with MCs are luminous GeV and weak TeV sources.
The \g-ray emitting isolated SNRs (Vela Jr \cite{velaH,velaF}, HESS J1731--347 \cite{j1731H}, RCW 86 \cite{rcwH,rcwF}, RX J1713.7--3946 \cite{j1713H,j1713F}, SN1006 \cite{sn1006H,Ara}, Tycho \cite{tycV,tycF}, Cas A \cite{casV,casF}) are all young SNRs (Vela Jr is $3000\pm1300$ yr old, but note that the age of HESS 1731--347 is weakly constrained). The SN 1006 upper limit (UL) on the 1-100\,GeV flux was derived by Araya $\&$ Frutos\,(2012)~\cite{Ara} assuming a point-like source at the position of SN 1006. The RCW 86 UL is extracted from Lemoine-Goumard et al.\,(2012) \cite{rcwF}. For HESS J1731--347 and HESS J1832--093, the flux of the closest Fermi detection is used as an UL (2FGL J1730.5--3350 and 2FGL J1827.4--0846 respectively) . 
The value of the GeV-TeV integrated flux ratio for isolated SNRs is much lower than for SNR-MC systems ($I_{GeV}/I_{TeV} \le 6\cdot10^{3}$) and characterizes hard GeV and strong TeV sources. 
Some evident differences appear between the MC-interacting and isolated SNRs. Sources for which interaction with a MC is possible lie between the two distinct populations of SNRs.

\vspace{0.8cm}
The inauguration of the fifth H.E.S.S. telescope last summer marks an encouraging step forward to disentangle features of confused sources, thanks to its improved sensitivity and resolution. The expected lower energy threshold of the instrument (E$\ge$10\,GeV) is a crucial tool for SNRs detected at GeV energies but not seen at TeV energies like W44. The detection of this strong GeV source resulting from the interaction of a SNR with a MC is still missing at VHE suggesting a cutoff or a break at a few  $\sim$\,10\,GeV.

\begin{small}

\section*{Acknowledgments}

See standard acknowledgements in H.E.S.S. papers not reproduced here due to lack of space.

\section*{References}
\begin{multicols}{2}

\end{multicols}

\end{small}

\end{document}